\documentstyle[aps]{revtex}
\tightenlines
\newcommand{\pt}{\partial}
 
\begin{document}
\title{\bf A Non-equilibrium Thermodynamic Framework
for the Dynamics and Stability of Ecosystems}
\author{K. Michaelian}
\address{Instituto de F\'isica, Universidad Nacional Aut\'onoma
de M\'exico \\ A.P. 20-364, 01000 M\'exico D.F., Mexico.}
\maketitle
\begin{abstract}
The population dynamics and stability of ecosystems of interacting 
species is studied from the perspective of non-equilibrium
thermodynamics by assuming that species, through their biotic and abiotic 
interactions, are units of entropy production and exchange in an open 
thermodynamic system with constant external constraints. Within the context 
of the linear theory of irreversible thermodynamics,
such a system will naturally evolve towards a stable {\em stationary} 
state in which the production of entropy within the ecosystem is at 
a local minimum value. It is shown that this extremal condition leads to 
equations for the stationary (steady) state population dynamics of 
interacting species, more general than those of Lotka-Volterra,
and to conditions on the parameters of the community interaction matrix 
guaranteeing ecosystem stability. The paradoxical stability of real 
complex ecosystems thus has a simple explanation within the proposed 
framework.
Furthermore, it is shown that the second law of thermodynamics 
constrains the inter- and intra-species interaction coefficients in the
sense of maintaining stability during evolution from one stationary state 
to another. A firm connection is thus established between the second law
of thermodynamics and natural selection.
\end{abstract}
\section{Introduction}
Ecosystems are complex. A typical ecosystem contains over 3000 species,
from bacteria to insects, plants and higher animals \cite{polis}. 
This size, coupled
with strong interactions among species and between species and the 
abiotic environment, leads to very complex population dynamics. 
Understanding the dynamics and the inherent stability of ecosystems is, 
however, of crucial importance in guiding wildlife management programs,
and in forecasting ecological catastrophes. Modeling of the population 
dynamics in the traditional  ecological framework
has been based on {\em ad hoc} extensions of Lotka-Volterra type 
equations \cite{may,may0,may2,mccann}. Apart from providing
little theoretical insight or empirical predictive power, such a 
framework implicitly contains a celebrated paradox concerning the 
improbability of stable, complex ecosystems \cite{may}.

This paper presents a new look at the questions of ecosystem stability and
dynamics from the perspective of linear irreversible thermodynamics (LIT). The
need to frame ecology within a thermodynamic paradigm 
has been recognized before \cite{wicken,wicken2,kay,schneider}. 
In the present work, the ecosystem is modeled as an open 
thermodynamic system over which a constant
free energy flow is impressed, sunlight.
Free energy also enters the ecosystem in 
the form of chemical potential, such
as nutrients. Interactions internal to the ecosystem, between the 
individuals of the species, and interactions between the individuals and 
the external abiotic environment, cause a time change of the total
entropy of the system. Assuming that the external constraints,
energy and nutrient flows, over the system
are constant, the linear theory of irreversible thermodynamics 
predicts \cite{prigogine}, and empirical results suggest, that the 
system will evolve towards a {\em stationary state} in 
which the local state variables (and hence global, extensive variables
such as the entropy $S$)
are constant in time. This stationary state is locally stable in the
sense that small fluctuations are naturally damped by flows generated
in the directions of the perturbing forces \cite{prigogine}.

Here, I show that writing the entropy change in time of an ecosystem as a 
many-body expansion in the interactions between individuals, and 
between individuals and the abiotic environment, and assuming constant
external constraints and the eventual establishment of a thermodynamic 
stationary state, 
leads to general dynamical equations for the populations of the
interacting species. A LIT condition of minimal entropy production in the
stationary state dictates conditions on the interaction 
parameters which assure the stability of the ecosystem. Furthermore, 
it is shown that
the second law of thermodynamics imposes restrictions on the inter-
and intra-specific interaction parameters, ensuring stability during the 
evolution of the system from one stationary state to another after the 
external constraints are changed, or after the system is significantly
perturbed.

\section{The Traditional Ecological Framework}
Population modeling in the traditional ecological framework is based on 
the equations \cite{may0},
\begin{eqnarray}
{dp_i(t) \over dt} = F_i(p_1(t), p_2(t), ... p_n(t)),
\label{eq:lkvt}
\end{eqnarray}
where $F_i$ is, in general, some empirically inspired, 
nonlinear function of the populations $p_i$
of the  $n$ species. For example, for the popular Lotka-Volterra equations 
(which have the stability characteristics of a much wider class of ecological
models employed in the literature \cite{may0}), $F$ takes the following form,
\begin{eqnarray}
F_i = p_i(b_i + \sum_{j=1}^n p_j c_{ij}).
\label{eq:lvol}
\end{eqnarray}
Of much interest in ecology, because of its frequent occurrence in nature, 
is the so called ecological {\em steady state} in which
all growth rates are zero, giving the fixed point, or steady state,
populations $p^*_i$,
\begin{eqnarray}
0 = F_i(p^*_1(t), p^*_2(t), ... p^*_n(t)).
\label{eq:eqpop}
\end{eqnarray}
The local population dynamics and stability in the neighborhood of the
fixed point can be determined by expanding Eqn. (\ref{eq:lkvt}) 
in a Taylor series about the steady state populations,
\begin{eqnarray}
{dx_i(t) \over dt} = F_i|_* + \sum_{j=1}^n \left[\left. \left.{\pt F_i \over 
\pt p_j}\right|_* x_j(t)
+ {1 \over 2}\sum_{k=1}^n \left[{\pt^2 F_i \over \pt p_j \pt p_k}\right|_* x_j x_k + 
... \right. \right. ,
\label{eq:taylor}
\end{eqnarray}
where $x_i(t) = p_i(t)-p_i^*$ and the $*$ denotes evaluation at the 
steady state. Since $F_i|_* = 0$, and close to the steady state
the $x_i$ are small, only the second term in 
the expansion (\ref{eq:taylor}) need be 
considered. In matrix notation, this gives,
\begin{eqnarray}
{\bf \dot{x}}(t) = {\bf A x}(t),
\label{eq:matrixeq}
\end{eqnarray}
where ${\bf x}(t)$ is a $n \times 1$ column vector of the population
deviations from steady state values, and the so called ``community matrix'' 
$\bf A$ has the components
\begin{eqnarray}
a_{ij} = \left.{ \pt F_i \over \pt p_j}\right|_* .
\label{eq:comp}
\end{eqnarray}
which represent the effect of species $j$ on the rate of change of 
population $i$ near the steady state.

The solution of equation (\ref{eq:matrixeq}) is 
\begin{eqnarray}
x_i(t) = \sum_{j=1}^n C_{ij} \exp(\lambda_j t)
\label{eq:soln}
\end{eqnarray}
where $\lambda_j$ are the eigenvalues of the matrix $\bf A$ and the 
integration constants $C_{ij}$ are determined from the initial
conditions. 

From equation (\ref{eq:soln}) it is obvious that local asymptotic stability 
near the steady state requires that 
the real parts of all the eigenvalues of $\bf A$ must be 
negative. This condition gives rise to very restrictive relations 
among the 
components $a_{ij}$ of the community matrix $\bf A$ \cite{may0}. For example,
it can be shown that for a $n=2$ species community it requires that
\begin{eqnarray}
a_{11}+a_{22} < 0, \nonumber \\
\end{eqnarray}
and 
\begin{eqnarray}
a_{11}a_{22} > a_{12}a_{21}.
\label{eq:conds}
\end{eqnarray}
For the Lotka-Volterra equations, Eqn. (\ref{eq:lvol}) this implies
\begin{eqnarray}
p_1^*c_{11} + p_2^*c_{22} < 0,
\label{eq:condslv1}
\end{eqnarray}
and
\begin{eqnarray}
c_{11}c_{22} > c_{12}c_{21}.
\label{eq:condslv2}
\end{eqnarray}
For a community of arbitrary $n$ species, it can be shown that the requirement 
that the $n\times n$
matrix $\bf A$ have all real parts of its eigenvalues negative is 
equivalent to the demonstration of the existence of a positive definite 
quadratic function $V = {\bf x^T P x}$ (Lyapunov function) 
having its derivative with respect to time
negative definite \cite{barnett}.

The restrictions on the components of the
community matrix for ensuring stability are thus specific,
and are more specific the more complex the ecosystem
\cite{may0}. Consequently, the probability that a randomly 
constructed community
will be stable decreases rapidly with the size of the ecosystem,
becoming practically zero at an ecosystem size of only
about 10 strongly interacting species \cite{may,gardner,yodzis}. This leads 
to a celebrated paradox: Without a mechanism for fine tuning the 
community matrix,
there should be little probability of finding stable complex
ecosystems. However, in nature, most ecosystems are very complex and most 
are stable \cite{polis,goldwasser}.

There have been many attempts to reconcile the theory with the field data
\cite{may0,mccann,mcnaughton,mccann0,roberts,deruiter,polis2,berlow}. 
The most plausible of these has been to invoke natural selection as the
mechanism for tuning the parameters of the community matrix 
\cite{may0}. This explanation, however, may be criticized as
being  tautological since there is no physical reason postulated 
for the selection of interaction coefficients leading to stability. 
Or, from another perspective, it leads to the celebrated problem of 
natural selection working on the evolution of a system of a 
population of one \cite{swenson}. A scenario in
which the elements of the community matrix are fortuitously chosen 
at random can be
discarded on the basis of statistical improbability of achieving 
stability for these large systems containing upwards of 3000 species.

\section{A Proposed Thermodynamic Framework}
The linear theory of irreversible
thermodynamics provides an interesting framework for accommodating
the problem of ecosystem dynamics and stability. The objective of
this paper is to demonstrate this by showing that steady state ecosystems 
have the signatures
of thermodynamic {\em stationary states}. The starting postulate of this
paper is that the total change of entropy of the ecosystem may be
written as a many-body expansion of entropy changes due to 
interactions among individuals. Specifically,
\begin{eqnarray}
{dS \over dt} = \sum_{i=1}^n \left[p_i\Gamma_{i} + \sum_{j=1}^n 
p_ip_j\Gamma_{ij} + \sum_{j,k=1}^n p_ip_jp_k\Gamma_{ijk} + 
O(4)\right].
\label{eq:dissfe}
\end{eqnarray}
The $\Gamma_i$ represent the change of entropy due to 1-body interactions 
of individuals with their abiotic environment (eg. evapotranspiration,
photo-synthesis, respiration, metabolic heat transfer to environment, 
etc.); $\Gamma_{ij}$ represents 2-body 
interactions between individuals (eg. predator-prey, competition, 
symbiosis, mutualism, etc.);
 $\Gamma_{ijk}$ correspond to the 3-body interactions, and
$O(4)$ represents 4-body and higher order interactions (eg. those required
for the functioning of
societies). Although this formulation of the total entropy change
is perhaps not the most general imaginable, it is a most common scheme 
chosen for 
systems in which the interacting constituents cannot be considered as 
ideal points in space-time, and in which no singularities are expected. 
Similar many-body expansion are used, for example,
for representing the interactions between extended, deform-able 
charged objects such as atoms in molecules and  clusters \cite{cao}, 
and nucleons in nuclei \cite{knutson}.

The total time change of entropy is a sum of an external term of no
definite sign, and, as required by the second law of thermodynamics, 
an internal term of positive definite sign,
\begin{eqnarray}
{dS \over dt} = {d_eS \over dt} + {d_iS \over dt}.
\end{eqnarray}
The external part of the change of entropy can be associated
with the one body interactions of the
individuals with their abiotic environment,
\begin{eqnarray}
{d_eS \over dt} =  \sum_{i=1}^n p_i\Gamma_{i}.
\label{eq:desdt}
\end{eqnarray}
The internal dissipative part is then associated with the 2-body and 
higher order interactions among the participating individuals,
\begin{eqnarray}
{d_iS \over dt} = \sum_{i=1}^n \left[\sum_{j=1}^n 
p_ip_j\Gamma_{ij} + \sum_{j,k=1}^n p_ip_jp_k\Gamma_{ijk} + 
O(4)\right] > 0.
\label{eq:disdt}
\end{eqnarray}
In the stationary state, ${dS / dt} = 0$, and since the internal dissipation
is positive by the second law, then, 
\begin{eqnarray}
{d_eS \over dt}  = \sum_{i=1}^n p_i \Gamma_i < 0,
\end{eqnarray}
indicating that at least one of the species 
must bring negative entropy into the
ecosystem, and that this negative entropy is greater than the positive 
entropy given back to the environment by the other one-body exchanges. 
This role is most often played by the photo-synthesizing species.

The inherent stability of a thermodynamic stationary state implies,
\begin{eqnarray}
{\pt \over \pt p_i} \left. \left[ {dS \over dt} \right]\right|_* = 0,
\label{eq:eqstab}
\end{eqnarray}
for all species $i$. The $*$ now denotes evaluation at the stationary state
populations. In the following, equation (\ref{eq:dissfe}) will
be truncated at the two-body terms. The justification for this is that, 
for most
ecosystems, higher order n-body interactions will be less probable
since they require n-body localization within a limited space-time volume.
The two-body truncation is in fact the norm in most ecological
studies \cite{may0,lassig,oliveira} with few exceptions \cite{oliveira2}.
This truncation, however, is certainly not valid for ecosystems 
with societal species, in which higher n-body interactions play an important
role. The more general dynamical equations and stability relations 
obtained from the complete equation
(\ref{eq:dissfe}) employing equation  (\ref{eq:eqstab}) 
will be discussed in a forthcoming article. Thus, taking
equation (\ref{eq:dissfe}) only to second order in the interactions,
Eqn. (\ref{eq:eqstab}) gives
\begin{eqnarray}
\Gamma_{i} + \sum_{j=1}^n p_j^*(\Gamma_{ij}+\Gamma_{ji}) = 0.
\end{eqnarray}
A simple change of variable makes these equations
recognizable as equivalents of those defining the steady state
populations in the ecological framework using the Lotka-Volterra 
equations, Eqs. (\ref{eq:lvol}), and conditions (\ref{eq:eqpop}). 
For example, for the case of $n=2$, the appropriate substitutions are,
$\Gamma_1 \equiv -b_1\sqrt{c_{21}/c_{12}}$, 
$\Gamma_{12}+\Gamma_{21} \equiv -\sqrt{c_{12}c_{21}}$ and 
$\Gamma_{11} \equiv -c_{11}\sqrt{c_{21}/c_{12}}/2$, with corresponding
definitions for $\Gamma_2$ and $\Gamma_{22}$.

In the stationary state, assuming linear phenomenological laws (see
below), the internal dissipation of entropy, 
$_i\dot{S} = d_iS/dt$ is a minimum \cite{prigogine}. In general, 
if $_i\dot{S}$ is a function of $n$ populations, the
condition for it to be a minimum is that the Hessian matrix,
\begin{eqnarray}
h_{ij} = \left.\left({\pt^2 _i\dot{S} \over \pt p_i \pt p_j} \right)\right|_*
\end{eqnarray}
is positive definite \cite{barnett}. As an example,
for $n=2$ species, $_i\dot{S}$ is a function of two variables,
$p_1$ and $p_2$, and the following two 
conditions must be satisfied \cite{swokowski};
\begin{eqnarray}
\left.{\pt^2_i \dot{S}  \over \pt {p_1}^2}\right|_{p_1^* p_2^*} > 0,
\nonumber\\
\left.{\pt^2_i \dot{S} \over \pt {p_1}^2} {\pt^2_i  
\dot{S} \over \pt {p_2}^2}\right|_{p_1^* p_2^*} - \left.
\left[ {\pt^2_i  \dot{S} \over \pt p_1 \pt p_2} \right]^2
\right|_{p_1^* p_2^*} > 0.
\end{eqnarray}
To second order in the interactions, this leads to the following 
conditions on the interaction parameters,
\begin{eqnarray}
\Gamma_{ii} > 0, \nonumber \\
4\Gamma_{11}\Gamma_{22} > (\Gamma_{12}+\Gamma_{21})^2.
\label{eq:conds2}
\end{eqnarray}
With the variable substitutions introduced above, these relations
can be recognized as sufficient conditions for stability of the steady state
populations in the ecological framework, equations (\ref{eq:condslv1})
and (\ref{eq:condslv2}).
That these conditions for arbitrary ecosystem size $n$ in this thermodynamic
framework are the same
as those imposed on the community matrix for stability in the ecological
framework can be 
demonstrated as follows: Consider the quadratic function
\begin{eqnarray}
V = {\bf x^T \Gamma x} = {\bf (p-p^*)^T \Gamma (p-p^*)}
\end{eqnarray}
where $\bf \Gamma$ is the matrix of entropy change due to 2-body interactions
$\Gamma_{ij}$.
The internal entropy production of the ecosystem at arbitrary populations 
$\bf p$, i.e. ${\bf p^T \Gamma p}$, and that at the stationary state 
populations, ${\bf {p^*}^T \Gamma p^*}$ are both positive 
definite by the second law of thermodynamics.
Since the internal production of entropy is at a minimum in the 
stationary state, $V$ is thus also positive definite. The time
derivative of $V$ is
\begin{eqnarray}
{d V \over dt}  = {d \left[{\bf (p-p^*)^T \Gamma (p-p^*)} \right] \over dt}.
\end{eqnarray}
A most general result of linear irreversible thermodynamics is that
the time change of the internal production of entropy 
\begin{eqnarray}
{d {\cal P} \over dt} = {d \over dt} \left[{d_iS \over dt}\right] =  
{d \left[{\bf p^T \Gamma p} \right] \over dt}
\end{eqnarray}
is negative semi-definite if the
external constraints are time-independent \cite{prigogine}. Since 
${d \left[{\bf p^T \Gamma p} \right] \over dt}$ has its maximum value of zero
at the stationary state populations $\bf p^*$, it is obvious that $dV \over dt$
is negative definite. We have thus found the Lyapunov function $V$
which establishes the local asymptotic stability of the community
matrix. An ecological steady state thus has the characteristics
of a thermodynamic stationary state and it is tempting to consider the
former as a particular case of the latter.

These stability conditions can be shown to be somewhat more general. For 
example,
consider the case of a system evolving from one stationary state to 
another \cite{prigogine}. The second law of thermodynamics requires that
always
\begin{eqnarray}
{d_iS \over dt} > 0,
\end{eqnarray}
or, to second order in the interactions,
\begin{eqnarray}
\sum_{i,j}^n p_ip_j \Gamma_{ij} > 0.
\end{eqnarray}
For example, for $n=2$ species 
\begin{eqnarray}
\Gamma_{11}p_1^2 + (\Gamma_{12}+\Gamma_{21})p_1p_2 + \Gamma_{22}p_2^2 > 0.
\label{eq:sqr}
\end{eqnarray}
Equation (\ref{eq:sqr}) can only always be satisfied, for whatever 
values of the populations, if the first of conditions (\ref{eq:conds2}) 
are met. For ecosystems in which $(\Gamma_{12} + \Gamma_{21})$ is negative,
the second of conditions (\ref{eq:conds2}) must also be met.
The second law of thermodynamics thus places restrictions on the values of
the inter- and intra-specific interaction parameters in the direction of 
securing ecosystem stability during evolution. The association of the 
second law with natural selection is thus implied.

\section{Phenomenological Laws and Reciprocity Relations}
The employment of the condition of minimal internal entropy production and 
that of the negative definiteness of the time change of the internal
entropy production implicitly
assumed the linearity of the phenomenological laws and the reciprocity
relations of Onsager \cite{prigogine}. To second
order in the interactions, the total change of entropy in the
ecosystem, Eqn. (\ref{eq:dissfe}), can be written in the form,
\begin{eqnarray}
{dS \over dt} = \sum_{i=1}^n \left[p_i\Gamma_{i} + \sum_{j=1}^n 
p_ip_j\left({\Gamma_{ij} + \Gamma_{ji}  \over 2}\right)\right].
\end{eqnarray}
In terms of generalized flows $J$ and forces $X$ \cite{prigogine},
\begin{eqnarray}
{d_iS \over dt} = \sum_i J_i X_i.
\end{eqnarray}
The flows and forces can thus be assigned in the following manner,
\begin{eqnarray}
J_i = \sum_j p_j 
\left({\Gamma_{ij} + \Gamma_{ji} \over 2}\right),
 \ \ X_i = p_i.
\end{eqnarray}
The generalized forces are thus the populations of the species and the
flows are the total changes of entropy due to the two-body interaction of
species $i$ with the rest of the species $j$.
The phenomenological relations are thus of the linear form,
\begin{eqnarray}
J_i = \sum_j L_{ij}X_j
\end{eqnarray}
where the phenomenological coefficients are,
\begin{eqnarray}
L_{ii} =& \Gamma_{ii} \nonumber \\ 
L_{ij} =& \left({\Gamma_{ij} + \Gamma_{ji} \over 2}\right).
\end{eqnarray}
From this and equation (\ref{eq:conds2}), or the condition following from
equation (\ref{eq:sqr}), it follows that,
\begin{eqnarray}
L_{ij} = L_{ji}, \ \ L_{ii} > 0.
\end{eqnarray}
The reciprocity relations of Onsager and the positive definite nature of
the proper phenomenological coefficients are thus satisfied to 2nd order
in the interactions, within or out of the stationary state.

\section{Discussion and Conclusions}
In the work presented here, interactions between the individuals have 
been taken only to second order. This was justified on the basis of
the smaller probability of higher n-body interactions, and was intended
for simplicity and for comparing results with traditional ecological 
approaches based on Lotka-Volterra type equations, which, in general, are
also of second order. Including higher order interactions means that
the phenomenological relations will then no longer be linear, implying that 
the condition of minimal entropy production no longer strictly applies. 
However, the more general result found by 
Prigogine and co-workers \cite{prigogine}, concerning the rate of internal 
entropy production, can still be used in this nonlinear regime. 
In a forthcoming paper it will be shown that this gives rise to wider 
spectrum of dynamical behavior for the populations.

In the case of changing external constraints,
or, more generally, an evolving ecosystem in which the phenomenological
coefficients (interaction parameters) cannot be treated as constants, 
again the linear
theory does not apply. However, it is still valid, as has been 
shown here, that the universal law of positive internal 
entropy production places restrictions on the possible values of the
interaction parameters in the direction of securing ecosystem stability.

Although we have shown in this paper that large, complex ecosystems are
constrained to stability by results from non-equilibrium thermodynamics, 
we have not argued why such systems might be favored over smaller, less
complex ones, as appears to be the case in nature. Although it is not
the intention of this paper to suggest a general evolutionary criterium
for ecosystems, a possible explanation, not in conflict with the
proposed framework, has been presented by Swenson \cite{swenson}.
Swenson argues that of all the possible paths available to a system after
the removal of an external constraint, a thermodynamic system will take
the path which increases the entropy of the system plus environment
at the fastest rate given the remaining constraints. Large, complex
ecosystems are more efficient at producing entropy than are smaller
ones, and thus would be favored by nature if this theory were correct. 

It is interesting that this apparent duality of ecosystems, to move 
towards stationary states of minimal entropy production over relatively
short time scales where the external constraints can be considered 
constant, and towards stable systems of higher internal entropy production
over longer evolutionary time scales, is mirrored within individuals.
It appears that an individual advances towards a state of minimal 
entropy production over development from birth to death \cite{prigogine},
while there is empirical evidence suggesting that there is an evolutionary 
trend in individuals towards higher metabolic rates (implying higher 
individual entropy production) \cite{zotin}.

In conclusion, non-equilibrium thermodynamics can serve as a useful
framework for describing the dynamics and stability of ecosystems.
In this framework, under the postulates of LIT, the stability of the community 
matrix is guaranteed, independent of its size, and there is thus 
no complexity-stability paradox. Under constant external constraints
the thermodynamic system evolves naturally towards a stable stationary state.
A stable stationary state, characterized by minimal internal entropy
production, implies a stable community matrix if the total change of
entropy of the ecosystem can be written as a many-body expansion of 
interactions between individuals as postulated here. The second law of 
thermodynamics places restrictions on the interaction parameters in the
sense of maintaining community stability 
during the evolution of the ecosystem from one stationary state to another.
This establishes a firm connection between natural selection and 
non-equilibrium thermodynamics and the second law of thermodynamics.

\acknowledgements{The author gratefully acknowledges useful comments on
the manuscript by J.L. Torres, J.A. Heras, J.M. Nieto, and 
L. Garcia-Colin Scherrer. The hospitality afforded by the Instituto 
de F\'{\i}sica y Matem\'aticas at the Universidad de Michoacana in 
Morelia, Mexico, and the financial support of CONACyT and DGAPA-UNAM 
are greatly appreciated.}

\end{document}